\documentclass[aps,prb,reprint,groupedaddress,showpacs,amsmath,amssymb,superscriptaddress]{revtex4-1}
\usepackage{subfigure}
\usepackage{graphicx}
\usepackage{dcolumn}
\usepackage{bm}

\usepackage[utf8]{inputenc}
\usepackage[T1]{fontenc}
\usepackage{mathptmx}

\usepackage{xcolor}

\begin{document}

\preprint{APS/123-QED}

\title{Growth and Characterization of Off-Stoichiometric LaVO$_3$ Thin Films}

\author{Biwen Zhang}
 \affiliation{Department of Physics, FSU, Tallahassee, FL 32306, USA}
 \affiliation{National High Magnetic Field Laboratory, FSU, Tallahassee, FL 32310, USA}
 
\author{Yan Xin}
 \affiliation{National High Magnetic Field Laboratory, FSU, Tallahassee, FL 32310, USA} 
 
 \author{Evguenia Karapetrova}
 \affiliation{Advanced Photon Source, Argonne National Laboratory, Argonne, Illinois, 60439, USA}
 
 \author{Jade Holleman}
 \affiliation{Department of Physics, FSU, Tallahassee, FL 32306, USA}
 \affiliation{National High Magnetic Field Laboratory, FSU, Tallahassee, FL 32310, USA}
 
 \author{Stephen A. McGill}
 \affiliation{National High Magnetic Field Laboratory, FSU, Tallahassee, FL 32310, USA} 

\author{Christianne Beekman}
 \email{beekman@magnet.fsu.edu}
 \affiliation{Department of Physics, FSU, Tallahassee, FL 32306, USA}
 \affiliation{National High Magnetic Field Laboratory, FSU, Tallahassee, FL 32310, USA}

\date{\today}

\begin{abstract}
LaVO$_3$ (LVO) has been proposed as a promising material for photovoltaics because its strongly correlated \textit{d} electrons can facilitate the creation of multiple electron-hole pairs per incoming photon, which would lead to increased device efficiency. In this study, we intentionally grow off-stoichiometric LVO films by changing the growth conditions such as laser fluence. Our aim is to study how deviating La:V stoichiometries affect the electronic properties of LVO thin films. We find that the off-stoichiometry clearly alters the physical properties of the films. Structural characterization shows that both La-rich and V-rich films have different levels of structural distortion, with La-rich (V-rich) films showing a larger (smaller) out-of-plane lattice parameter compared to what one would expect from epitaxial strain effects alone. Both types of films show deviation from the behavior of bulk LVO in optical measurement, i.e., they do not show signatures of the expected long range orbital order, which can be a result of the structural distortions or the presence of structural domains. In transport measurements, La-rich films display clear signatures of electronic phase separation accompanying a temperature induced metal-insulator transition, while V-rich films behave as Mott insulators. The out-of-plane lattice parameter plays a crucial role in determining the transport properties, as the crossover from Mott-insulating to disorder-induced phase-separated behavior occurs around a lattice parameter value of 3.96~\AA, quite different from what has been previously reported. 
\end{abstract}

\maketitle

\section{Introduction}
Strongly correlated electron systems have attracted intense research interests over decades due to their wide range of functionalities. Properties including colossal magneto-resistance\cite{CMR}, high-T$_C$ superconductivity\cite{Capone} and orbital ordered states\cite{OO-LSVO,OO-LVO} arise from the complicated nature of \textit{d} electrons in transition metals\cite{Mott1994MetalInsulatorTransitions}. Orthovanadates RVO$_3$ (R = rare earth or yttrium) are strongly correlated insulators (SCI) whose properties can be affected by the choice of the R elements\cite{Sage2007,Martinez-Lope2008EvolutionDatab}. 
LaVO$_3$ (LVO), as a narrow bandgap SCI, exhibits a variety of properties that are of both fundamental and technological interest. For example, Coulter et al.\cite{Coulter2015Photo-ExcitedInsulators} theoretically proposed that some SCI can become novel photovoltaic materials with enhanced solar cell efficiency, because the strong Coulomb interactions in the \textit{d} electrons in vanadium atoms can allow the material utilize the excess energy from a "hot" electron to create multiple electron-hole carriers through the process called Impact Ionization (II)\cite{Coulter2014OptoelectronicMaterials,Coulter2015Photo-ExcitedInsulators,Holleman2016EvidenceDioxide,Manousakis2010PhotovoltaicInsulators}. For LVO, the calculated II rate is of the order of 10\textsuperscript{15} s\textsuperscript{-1}, which is orders of magnitude faster than the rate of electron-phonon interaction 
($\sim$10\textsuperscript{12} s\textsuperscript{-1}) in this material\cite{Coulter2015Photo-ExcitedInsulators,Zhang2017High-QualityMaterial}; the bandgap of bulk LVO is 1.1 eV\cite{Wang2015DeviceMaterial,Zhang2017High-QualityMaterial}, which makes it theoretically ideal to realize the II process with solar energy. In addition to that, LVO also exhibits a variety of intriguing physical properties such as orbital-order\cite{Sage2007}, structural transitions \cite{Miyasaka2002AnisotropyYVO3}, metal-insulator transitions, and it shows promise as a basis for a p-type transparent conducting oxide \cite{Hu2020,Hu2018,Zhang2015}. Most importantly, numerous thin-film works have clearly shown that these functional properties can be tuned by applying external perturbations, such as applied current and structural distortions induced by epitaxial strain and doping \cite{Hotta2006GrowthFilms,Rotella2012OctahedralFilms, Rotella2015StructuralFilms, Zhang2015Self-regulatedEpitaxy, choi2010}. This makes LVO a promising candidate for strain- and doping-driven engineering of materials properties. 


\begin{figure}[b]
\centering
\includegraphics[width=3.4in]{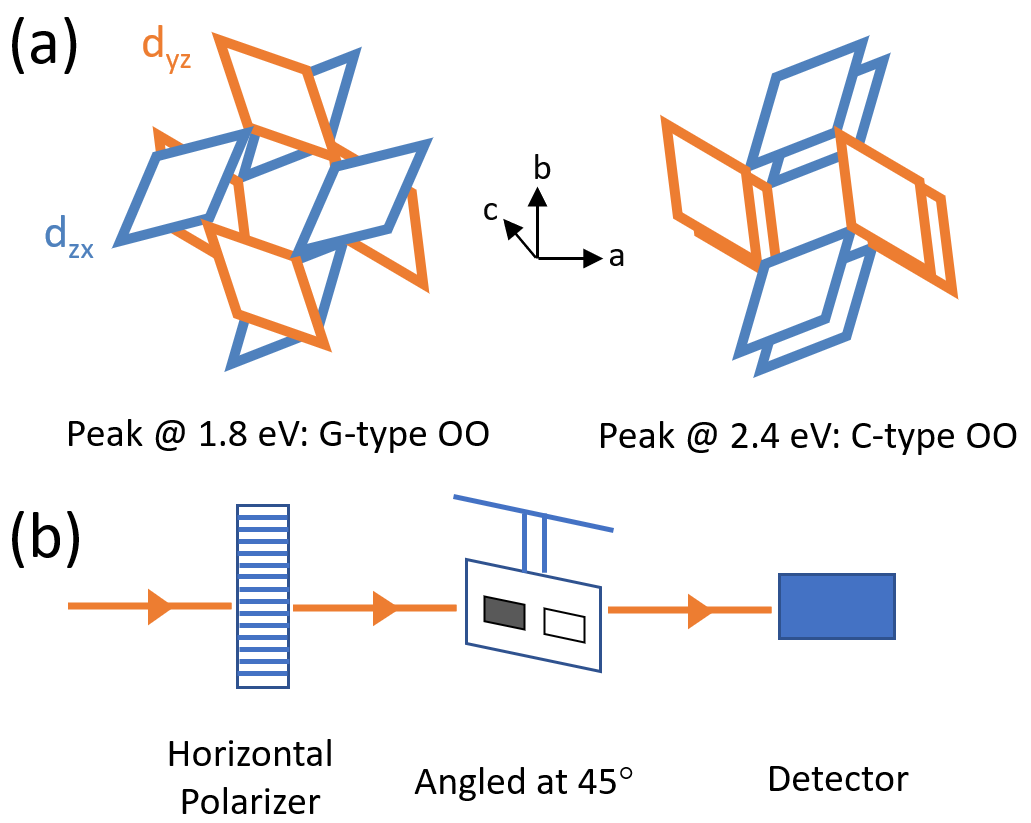}
\caption{\label{fig:Structure}(a) Two possible orbital configurations in the RVO$_3$ family. Left: G-type OO. right: C-type OO, which have corresponding peaks in absorption spectra at 1.8 and 2.4 eV, respectively. (b) Optical measurement setup to detect types of orbital ordering. A sample stage that is angled at 45$^\circ$ with regard to the incoming light is located inside of a cryostat. A horizontally polarized white light source is used to illuminate the samples. The transmitted light intensity is measured by a visible-near infrared CCD detector.}
\end{figure}

Bulk LVO is orthorhombic (\textit{Pbnm}) or pseudocubic at room temperature, with lattice constants a\textsubscript{p} $\approx$ b\textsubscript{p} $\approx$ c\textsubscript{p} $\approx$ 3.924~\AA \cite{Martinez-Lope2008EvolutionDatab,Rotella2012OctahedralFilms}. The level of distortion of a RVO$_3$ lattice can be evaluated by the rotation and tilting of the VO$_6$ octahedra or by the angle of V-O-V bonds, and such distortions will affect how the \textit{d} orbitals order inside vanadium atoms. Generally, for orthorhombic RVO$_3$, the \textit{d} orbitals adopt a so-called C-type orbital ordering (C-type OO) configuration (Fig.~\ref{fig:Structure}a right), where the \textit{d\textsubscript{yz}} and \textit{d\textsubscript{zx}} orbitals are aligned out of phase in the \textit{ab}-plane, but in phase along the \textit{c}-direction; if a RVO$_3$ has a more distorted monoclinic structure, an alternative G-type orbital ordering (G-type OO) (Fig.~\ref{fig:Structure}a left) can form, in this type of ordering, \textit{d\textsubscript{yz}} and \textit{d\textsubscript{zx}} are aligned out-of-phase in all directions \cite{Sage2006}. Among the RVO$_3$ family, LVO is the least distorted crystal and it has the largest V-O-V bonding angle \cite{Martinez-Lope2008EvolutionDatab,Sage2007}, but it undergoes a structural transition to a monoclinic phase when it is cooled to below 141~K \cite{Miyasaka2002AnisotropyYVO3}. As a result, the \textit{d} orbitals will order to the G-type OO state at low temperature. Previous reports showed that optical absorption is effective in detecting the structural phase transition associated with the formation of G-type OO in LVO\cite{Miyasaka2002AnisotropyYVO3,Tomimoto2003UltrafastLaVO3}. This is because the energies to excite electron hopping between neighboring vanadium sites are different in the C-type OO and G-type OO phases. In the case of bulk LVO, its C-type OO absorption peak is at 2.4~eV while the G-type OO peak is at 1.8~eV, respectively.

Here we present the study on using Pulsed Laser Deposition (PLD) to grow off-stoichiometric LaVO$_3$ thin films on SrTiO$_3$ (STO) (001) substrates. We investigate how the presence of La:V off-stoichiometry affects the physical properties of LVO thin films. In line with studies on other complex oxides (e.g. manganites \cite{CMR} and cuprates\cite{cuprates}), one expects to find rich physics upon straining, doping or distorting the structure of orthovanadate thin films as well.
According to existing literature, PLD is the most common technique to grow LVO films\cite{Hotta2006GrowthFilms,Rotella2015StructuralFilms,Wang2015DeviceMaterial,Meley-2018,Choi2000,Rotella2012OctahedralFilms}. Previous reports show that growth conditions can control the oxygen stoichiometry of the films, sometimes leading to the formation of unwanted LaVO$_4$ parasitic phases \cite{Hotta2006GrowthFilms,Wadati,Wang2015DeviceMaterial}. Some have investigated the optical and transport properties of PLD-grown LVO thin films, including ultrathin films \cite{Wang2015DeviceMaterial,Hotta2007PolarInterface}. However, these works only looked at stoichiometric thin films. Off-stoichiometric LVO thin films studies are all done on Molecular Beam Epitaxy (MBE) grown films, where the La:V stoichoiometry is shown to alter the structural properties of the films \cite{Zhang2015Self-regulatedEpitaxy,Zhang2017High-QualityMaterial}. In this work, we report on the use of PLD to grow off-stoichiometric LVO thin films by means of controlling laser fluence. 
Characterizations on the off-stoichiometric LVO films show that the lattice mismatch between the films and the substrate causes the films to be strained. 
From structural characterizations using x-ray diffraction (XRD) and transmission electron microscopy (TEM), we learn that the amount of elongation of the out-of-plane lattice constant depends on the La:V stoichiometry. The V-rich (La-rich) films have a smaller (larger) unit cell volume compared to the expected unit cell volume based on the reported Poisson ratio for bulk LVO and the epitaxial strain \cite{Masset,Brahlek}. Furthermore, the film unit cells are not single-oriented, similar to previous reports \cite{Masset}.
From absorption measurements (see Fig.~\ref{fig:Structure}b) we find the coexistence of C-type and G-type absorption peaks. We do not see significant changes in the spectral weight of G-type OO peak as the films are cooled down. This indicates that the long range G-type orbital ordering that was observed at low temperature in bulk crystals is absent in our thin films.
The transport measurements show a clear stoichiometry dependence. The V-rich films show correspondence to bulk LVO, but the La-rich films have unexpected metal-insulator transitions as well as non-Ohmic current-voltage curves pointing to electronic phase separation.
These observations clearly indicate that the electronic properties of off-stoichiometric thin films are largely affected by the out-of-plane lattice parameter, i.e., the amount of distortion to the LVO unit cell.

\section{EXPERIMENTAL}

LVO films are grown by the PLD technique, in which a KrF excimer laser ($\lambda = 248\ \textnormal{nm}$) is focused on a ceramic target LaV\textsubscript{1.2}O$_4$ (half of the La-rich films were grown using a LaV\textsubscript{1.1}O$_4$ target). The LaV\textsubscript{1.2}O$_4$ targets are prepared from La$_2$O$_3$ (99.9\% purity) and V$_2$O$_5$ (99.9\% purity) powders with the ratio of 1:1.2. Excess V$_2$O$_5$ powder is used to compensate for losses of the highly volatile vanadium element during the target sintering process (see Supplemental Material\cite{suppmat}  and ref. \cite{Kim2011Room} for more details on the target fabrication recipe). The STO subtrates are pre-treated using deionized-water in a leaching and thermal annealing method to obtain atomically flat and TiO$_2$ terminated surfaces\cite{Connell2012PreparationProcedure}. The laser fluence ranges from $\sim$0.6 J/cm\textsuperscript{2} to $\sim$2.0 J/cm\textsuperscript{2}, while the repetition rate of the laser is kept at 1 Hz. All films are deposited at the order of 10\textsuperscript{$-7$} mTorr base pressure to stabilize the perovskite phase of LaVO$_3$. The temperature of the substrate is kept in the range of 600-650$^\circ$C during growth\cite{Hotta2006GrowthFilms}.

\begin{figure*}[hbt!]
\includegraphics[width=6.4in]{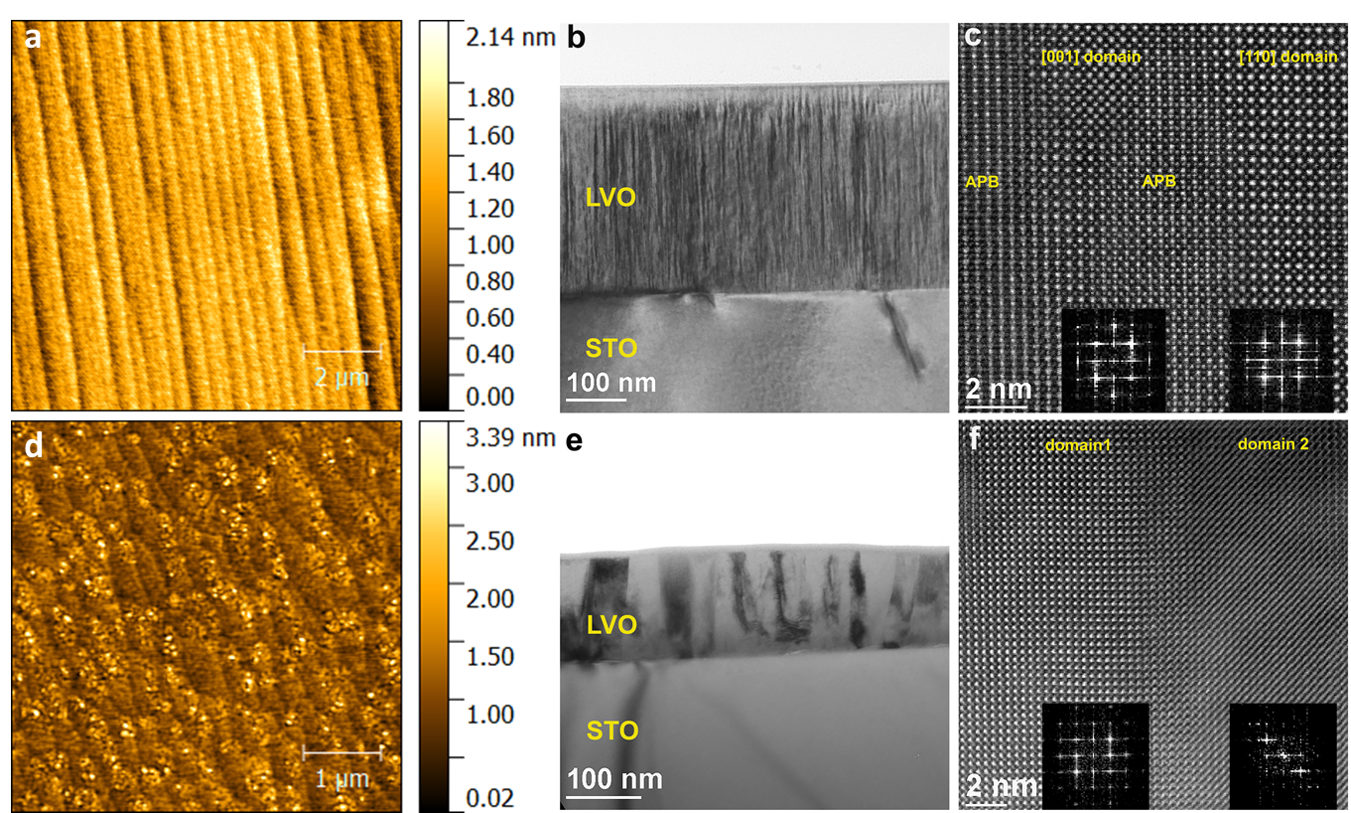}
\caption{\label{fig:AFM}AFM and TEM images for LVO films with different stoichiometry. Top row: V-rich film LVO-V1 (a) AFM image showing a surface with unit cell terraces (RMS roughness: 2.8 \AA). (b) Low magnification bright field TEM image of the cross-section view. The streaky lines inside the film are contrast from the anti-phase domains (c) Atomic resolution HAADF-STEM image. Insets: (left) Fast Fourier Transform (FFT) from [001] domain; (right) FFT from [110] domain.
Bottom row: La-rich film LVO-L1 (d) AFM image showing a surface with unit cell terraces (RMS roughness: 3.5 \AA). (e) Low magnification bright field TEM image of the cross-section view. (f) Atomic resolution HAADF-STEM image. Inset: FFTs from domain 1 (left) and domain 2 (right). } 
\end{figure*}

The morphology of the LVO films is characterized by atomic force microscopy (AFM) using an Asylum Research microscope. The structure is studied using a Scintag DMC-105 x-ray diffractometer, a Rigaku SmartLab SE x-ray diffractometer (both with Cu K$\alpha_1$, $\lambda$ = 0.154 nm), and the Advanced Photon Source at Argonne National Laboratory (photon energy = 12.398 keV, $\lambda$ = 1 \AA) at beamline 33-BM-C \cite{KARAPETROVA201152}. Detailed information on the microstructure and the chemical composition are investigated using a probe-aberration-corrected cold field emission JEM-ARM200cF at 200 kV, which is equipped with Oxford energy dispersive spectroscopy (EDS) detector for composition analysis. Electrical resistivity $\rho$(T) measurements are carried out from 300~K to 2~K using a Quantum Design physical property measurement system (PPMS) and Keithley external electronics model 2182 and model 6221. The samples for PPMS measurements are cut to strips (length: 5~mm, width: 0.5~-~1.2~mm) and we use a 4-point probe measurement geometry via evaporated gold contacts. The optical properties are characterized using absorption measurements (see Fig.~\ref{fig:Structure}b for a schematic of the setup). For the absorption measurements we used an Ocean Optics USB2000 Spectrometer with a QTH (quartz-tungsten-halogen) lamp as the source. The CCD detector is a Si-based photodiode array that is uncooled (integrated into the USB spectrometer by Ocean Optics). The resolution and range of the spectrometer are, 3 meV and 1.5 eV - 3.1 eV, which covers the range of our interest, the LVO \textit{d-d} transitions. The incoming light is polarized to the horizontal direction and the sample stage is angled at 45$^\circ$ with regard to it. 
An LVO film and a bare STO substrate are fixed to the sample stage in order to take their spectra at each temperature.  This setup is intentionally designed this way so that the light can interact with the film both along the \textit{c} axis and \textit{ab}-plane. The backside of all samples measured in this optical setup are hand-polished until optically smooth using a 3-micron diamond lapping film. The relative absorbance \textit{A}\textsubscript{$\lambda$} is calculated with:
\begin{equation}
    A_\lambda = -log_{10}\frac{LVO_\lambda-Dark_\lambda}{STO_\lambda-Dark_\lambda}
\end{equation}
where LVO\textsubscript{$\lambda$} and STO\textsubscript{$\lambda$} are the transmitted light intensity of LVO film and STO substrate, respectively. Dark\textsubscript{$\lambda$} is the dark current noise from the detector.   

\begin{figure*}[hbt!]
\includegraphics[width=7in]{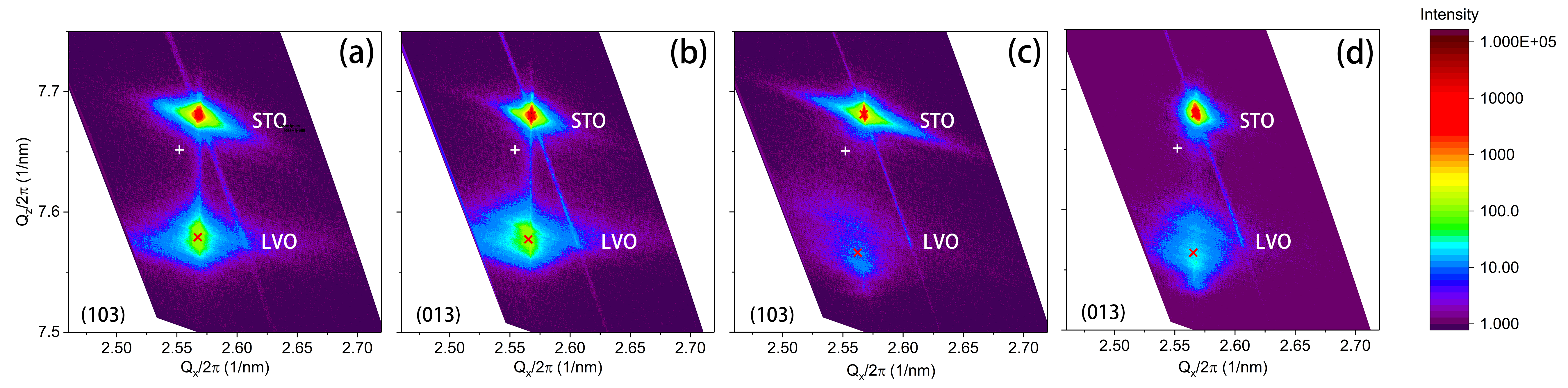}
\caption{\label{fig:RSM}Reciprocal space mapping for LVO-V2 and LVO-L1. Panel (a) and (b) show the RSM for LVO-V2 around its (103) and (013) peaks, respectively. Panel (c) and (d) shows the RSM for LVO-L1 around its (103) and (013) peaks, respectively. The STO and LVO peaks are labeled in each panel. The white cross-hairs indicate the peak position for stoichiometric bulk LVO. The red "x" marks the position of maximum intensity for each film peak.}
\end{figure*}

\section{RESULTS AND DISCUSSION}

\subsection{Morphology and Structure}
First, we discuss the influence of the stoichiometry on the morphology and microstructure of the thin films. The results from AFM measurements show that the surface morphology of the LVO films is not very sensitive to stoichiometry. AFM images in Fig~\ref{fig:AFM}(a) and \ref{fig:AFM}(d) show that both types of the films have unit-cell terraces with comparable RMS roughness values, 2.8 \AA~for LVO-V1 and 3.5 \AA~for LVO-L1.
The microstructure of the film does depend on stoichiometry, as evidenced by the TEM images in Fig. \ref{fig:AFM}. The La:V stoichiometry of the films was determined using the EDS and/or the growth conditions used and is 45:55 for the V-rich and 55:45 for the La-rich films. The cross-section view of the V-rich films [Fig.~\ref{fig:AFM}(b) and \ref{fig:AFM}(c)] revealed that there are high density antiphase nanodomains with different c-axis orientation relative to the substrate. The orientations of these domains are determined to be [001] (\textit{c}-axis pointing to the viewer) and [110] (\textit{c}-axis pointing to the top surface) by doing Fourier transformation of the images \cite{Rotella2015StructuralFilms,Vrejoiu2016ProbingScattering}. 
However, the La-rich film shows different microstructures, and it has lower density of antiphase domains, but larger grains with tilted c-axis forming low angle grain boundaries instead [see Fig.~\ref{fig:AFM}(e) and \ref{fig:AFM}(f)]. The more random tilts of the c-axis in the La-rich films is in line with the reduced structural coherence observed in these films in the RSM measurements, as discussed next. 

\begin{figure}[t]
\includegraphics[width = 3.4in]{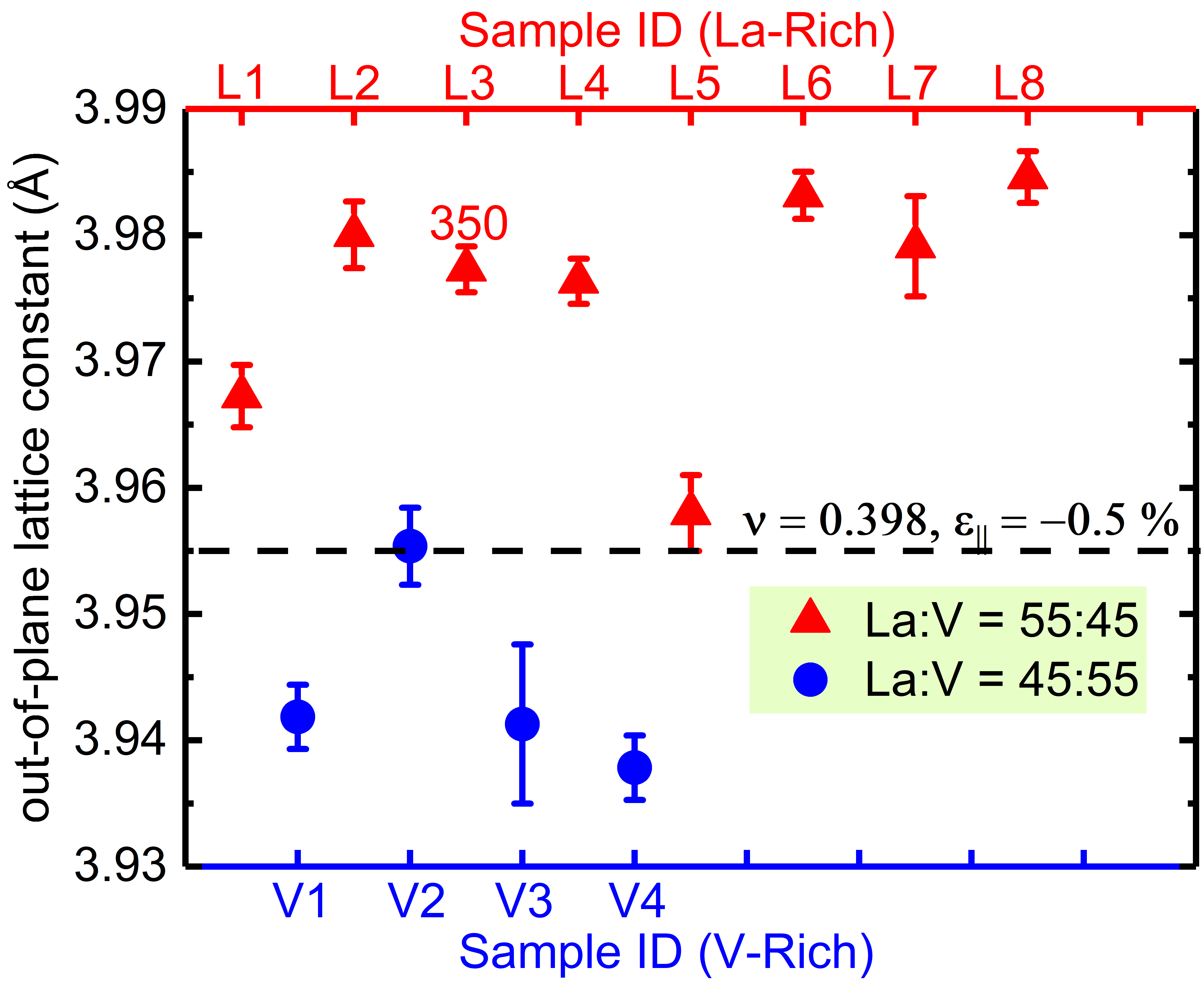}
\caption{\label{XRD_c} Summary of the out-of-plane lattice constants for the films discussed in the manuscript. Each film was grown using nominally the same growth conditions, except for the laser fluence, which was set to $J$ $\sim$0.6 J/cm$^2$ and $J$ $\sim$ 1 J/cm$^2$ (or higher), for La-rich and V-rich films, respectively. The La:V stoichiometry was determined using EDS (La:V = 55:45 and La:V = 45:55) or inferred from the growth conditions. The film thickness was around 200~nm unless indicated otherwise, i.e., L3: 350~nm, as determined from TEM and from the number of pulses used during growth. The dashed line is the expected lattice parameter based on a Poisson ratio of $\nu$ =0.398 \cite{Masset,Brahlek} and epitaxial strain of $\epsilon_{||}$ = -0.5 $\%$.}
\end{figure}



\begin{figure}[h]
\centering
\includegraphics[width=3.4in]{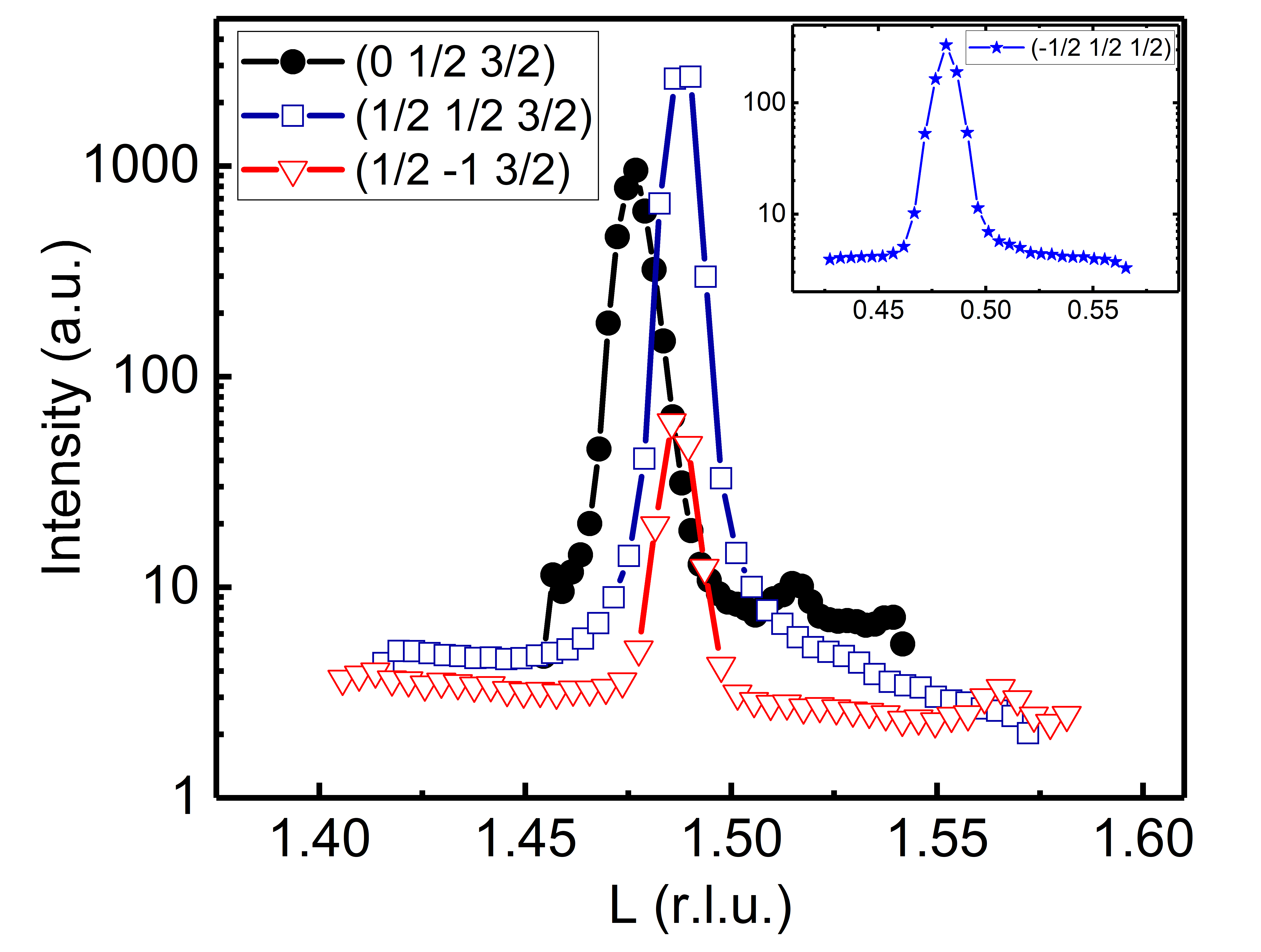}
\caption{\label{fig:Half} Half order peaks of LVO-L1. With the diffraction peaks shown, we are able to determine an \textit{a\textsuperscript{+}a\textsuperscript{-}c\textsuperscript{-}} tilting pattern in our films.}
\end{figure}

Representative reciprocal space maps (RSMs) on a V-rich film (LVO-V2) and a La-rich film (LVO-L1) reveal that our films are strained onto the STO substrates (Fig.~\ref{fig:RSM}). The lattice constants extracted from these RSMs are a\textsubscript{p} = 3.908(1) \AA, b\textsubscript{p} = 3.905(1) \AA ~and c\textsubscript{p} = 3.955(2) \AA ~for LVO-V2, and a\textsubscript{p} = 3.907(1) \AA, b\textsubscript{p} = 3.911(1) \AA ~and c\textsubscript{p} = 3.969(4) \AA ~for LVO-L1. (see Supplemental Material\cite{suppmat} for how the lattice constants are extracted and calculated.) The contraction in the \textit{a} and \textit{b} axes and the elongation in the \textit{c} axis (out-of-plane direction) in these films are expected because there is a compressive in-planar strain of about 0.5$\%$ induced by the lattice mismatch between STO (3.905 \AA) \cite{Navi2012ThermochemistrySolutions} and LVO. Furthermore, LVO-L1 has weaker film peak intensity than LVO-V2. Since the two films are similar in thickness (around 200 nm), such a difference indicates that the structure in the La-rich film is less coherent, which is consistent with the structural domains observed in TEM (see Fig. \ref{fig:AFM}). Apart from the change in the dimensions of the LVO unit cell due to the epitaxial strain effect, we notice that our LVO films with different La:V stoichiometry have differences in their out-of-plane lattice constants. Fig. \ref{XRD_c} summarizes the out-of-plane lattice constant \textit{c} extracted from $\theta-2\theta$ scans on our films (see Supplemental Material \cite{suppmat}). The \textit{c} parameter for the La-rich films is approximately 1\% greater than for the V-rich films. Such a difference is subtle but consistent, and in agreement with previous studies on the non-stoichiometric bulk LVO samples\cite{Seim1998NonStoichiometricLI,Gharetape2011EffectLaVO3}.




To further characterize the lattice structure in our films, we performed synchrotron x-ray diffraction measurements on LVO-L1 and LVO-V2. From the scans, we obtain half order peaks (see Fig.~\ref{fig:Half}) such as even-odd-odd ([$0\frac{1}{2}\frac{3}{2}$]), odd-even-odd ([$\frac{1}{2}0\frac{3}{2}$]) and odd-odd-odd ([$\frac{1}{2}\frac{1}{2}\frac{3}{2}$], [-$\frac{1}{2}$$\frac{1}{2}$$\frac{1}{2}$]). Based on these observed peaks, we conclude an \textit{a\textsuperscript{+}a\textsuperscript{-}c\textsuperscript{-}} octahedral tilting pattern in both films. This is in agreement with the pattern reported by Rotella et al \cite{Rotella2012OctahedralFilms} (\textit{a\textsuperscript{-}a\textsuperscript{+}c\textsuperscript{-}}) in their stoichiometric LVO thin films.  

\subsection{Absorption}

\begin{figure}[h]
\centering
\includegraphics[width=3.2in]{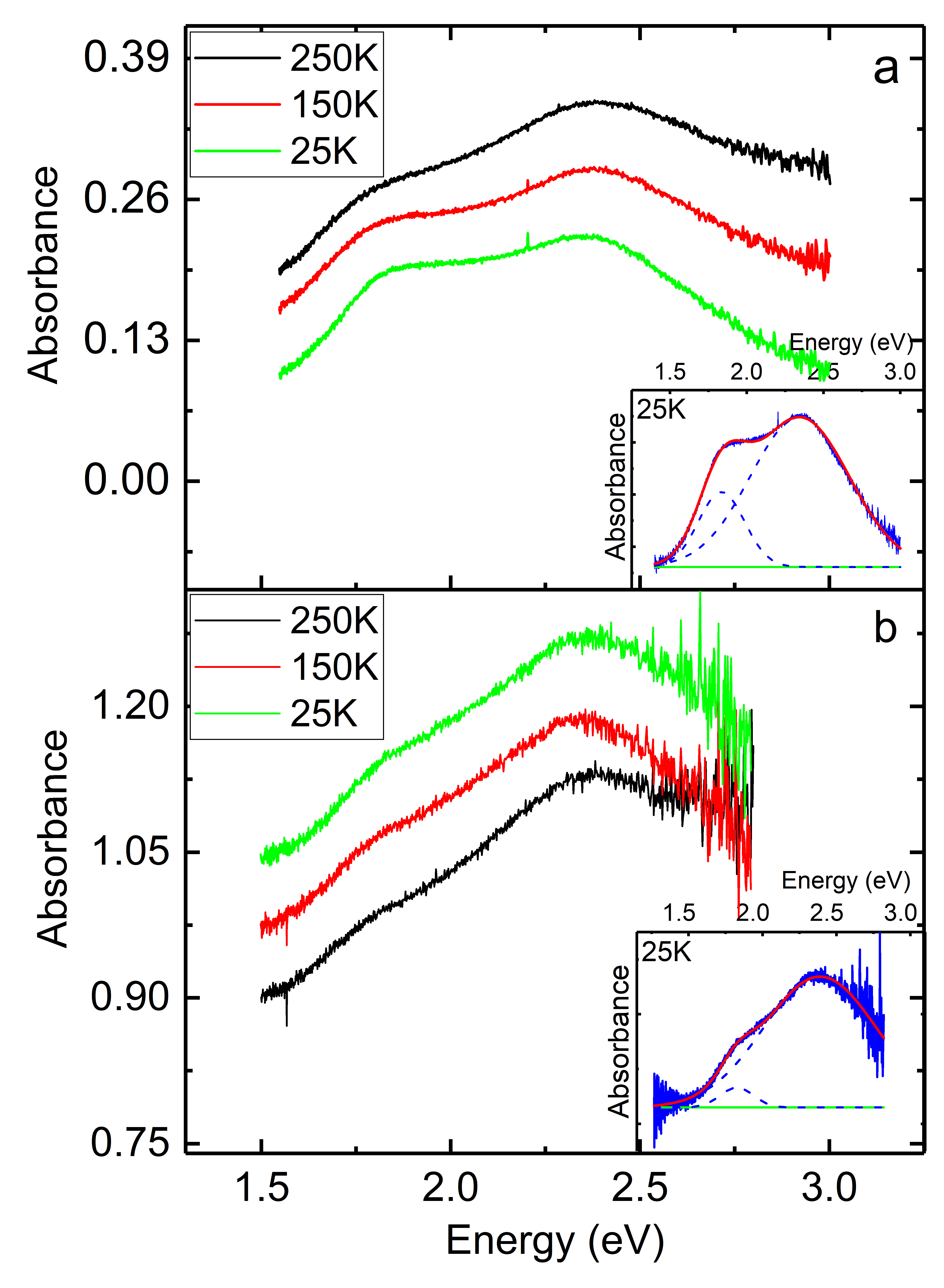}
\caption{\label{Abs_T}Temperature-dependent absorption spectra on (a) LVO-L2 (integration time: 250~ms) and (b) LVO-V2 (integration time: 400 ms) at 250~K, 150~K and 25~K. 
Insets are Gaussian fittings of the 25~K spectra. }
\end{figure}

\begin{table*}[hbt!]
\caption{\label{Abs_tab}Peak analysis of absorption spectra at different temperatures.}
\begin{ruledtabular}
\begin{tabular}{cccccccccccccc}
& &\multicolumn{4}{c}{LVO-L2}&\multicolumn{4}{c}{LVO-V2}\\
&Temperature &\multicolumn{2}{c}{G-type OO} &\multicolumn{2}{c}{C-type OO}
&\multicolumn{2}{c}{G-type OO} &\multicolumn{2}{c}{C-type OO}\\
& &Energy(eV) &\%Weight &Energy(eV) &\%Weight &Energy(eV) &\%Weight &Energy(eV) &\%Weight\\\hline
&250 K &1.78 &3.73 &2.40 &96.27 &1.77 &3.04 &2.39 &96.96\\
&200 K &1.79 &6.21 &2.40 &93.79 &1.78 &4.86 &2.39 &95.14\\
&150 K &1.80 &10.01 &2.39 &89.99 &1.80 &6.54 &2.35 &93.46\\
&100 K &1.81 &11.05 &2.35 &88.95 &1.82 &5.63 &2.39 &94.37\\
&25 K &1.83 &15.80 &2.33 &84.20 &1.82 &4.88 &2.39 &95.12\\
\end{tabular}
\end{ruledtabular}
\end{table*}



Temperature dependent absorbance is done on various films and representative spectra (for LVO-L2 and LVO-V2) at selected temperatures are shown in Figure \ref{Abs_T}. To analyze the absorption peaks, spectra are fitted to Gaussian function with the fitting background set to be a constant value (the minimum of each spectrum) (Fig.~\ref{Abs_T} insets).
The fit results can be found in table \ref{Abs_tab}. From figure \ref{Abs_T} and table \ref{Abs_tab}, we can see both films have a temperature dependence when being cooled down. The G-type OO absorption peak slowly gains spectral weight, but not as drastic as what others reported in bulk LVO in the low temperature regime\cite{Tomimoto2003UltrafastLaVO3,Miyasaka2002AnisotropyYVO3}. The energies for G-type and C-type OO peaks in both types of films are similar to bulk values. Though there seems to be some temperature dependent energy shifts, these are more likely due to fitting error because other absorption peaks outside our measurement range (i.e., transitions between the O 2p and V 3d bands) can contribute to the spectrum as a temperature dependent "background"\cite{Wang2015DeviceMaterial}. From these measurements it is clear that our LVO-L2 and LVO-V2, do not display clear evidence of OO at low temperatures, but the two peaks of \textit{d-d} transitions are distinguishable (see table \ref{Abs_tab}) at all temperatures. Here we note that Zhang et al.\cite{Zhang2017High-QualityMaterial} reported room-temperature absorption spectra on stoichiometric and off-stoichiometric LVO thin films grown using MBE. They found that both the stoichiometric and the La-rich films kept a pronounced two-peak feature (peaks at 1.7 and 2.2 eV, respectively), but for the V-rich films, the two peaks wash out with increasing V-concentration. Hence, our films may be closer to optimal stoichiometry in comparison and the optical properties appear to be quite insensitive to changes in stoichiometry of the order of $\sim 5\%$.

Furthermore, the apparent absence of orbital order in our films at low temperature has multiple possible causes. The presence of structural domains as seen in our TEM images will hinder formation of long-range orbital order, even if the individual domains are ordered. It is also possible that the extension of the out-of-plane direction plays a role in suppressing the formation of orbital order. The excess or deficiency of the vanadium alters the V-O-V bond lengths and angles\cite{Gharetape2011EffectLaVO3,Seim1998NonStoichiometricLI}, leading to a more distorted bonding scheme possibly disrupting orbital ordering.

\subsection{Resistivity}

\begin{figure}[h]
\centering
\includegraphics[width=3.3in]{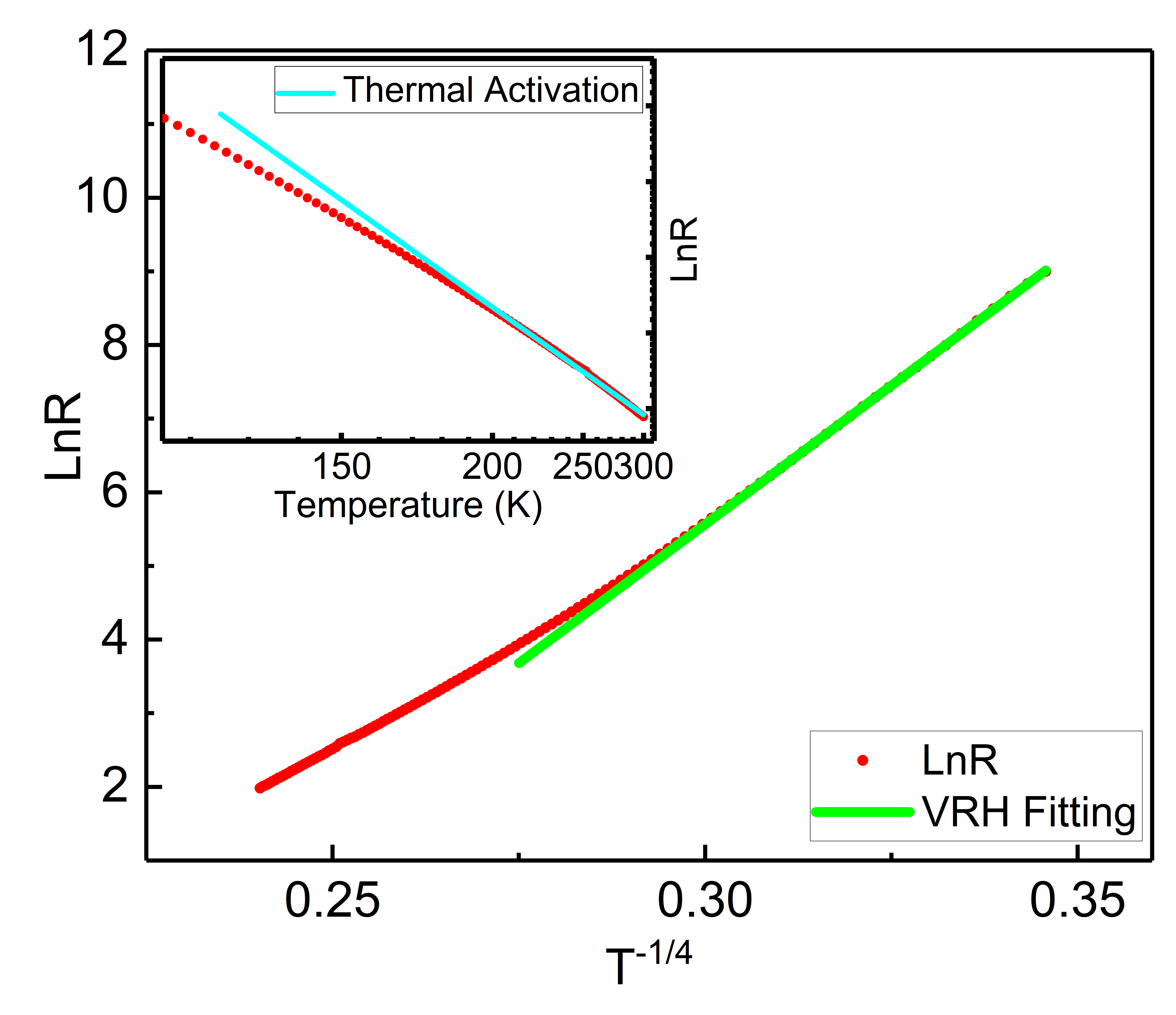}
\caption{\label{PV} Resistance measurement on a representative V-rich film LVO-V2 measured with 0.001 mA, shows that its low temperature resistance can be well fitted to the Mott variable range hopping model of a three dimensional material ($\sim$ T\textsuperscript{-1/4}), the resistance at high temperature can be explained using thermal activation model (see inset).}
\end{figure}

Bulk LVO is a Mott insulator \cite{Miyasaka2000CriticalLa1-xSrxVO3,PhysRevB.98.075124}.
All V-rich films described in this paper behave as Mott insulators. Figure \ref{PV}a plots the temperature dependent resistance of a representative V-rich film (LVO-V2): 
its high temperature resistance can be fitted with thermal activation model with an activation energy of 73.3 meV (Fig.~\ref{PV}a inset); in the low T region, the resistance can be explained using the Mott variable range hopping model of a three dimensional material\cite{MOTT19681,Efros1975CoulombSystems}. All V-rich films indicated in Fig. \ref{XRD_c} with the blue symbols, show this behavior. 

\begin{figure}[h]
\centering
\includegraphics[width=3.3in]{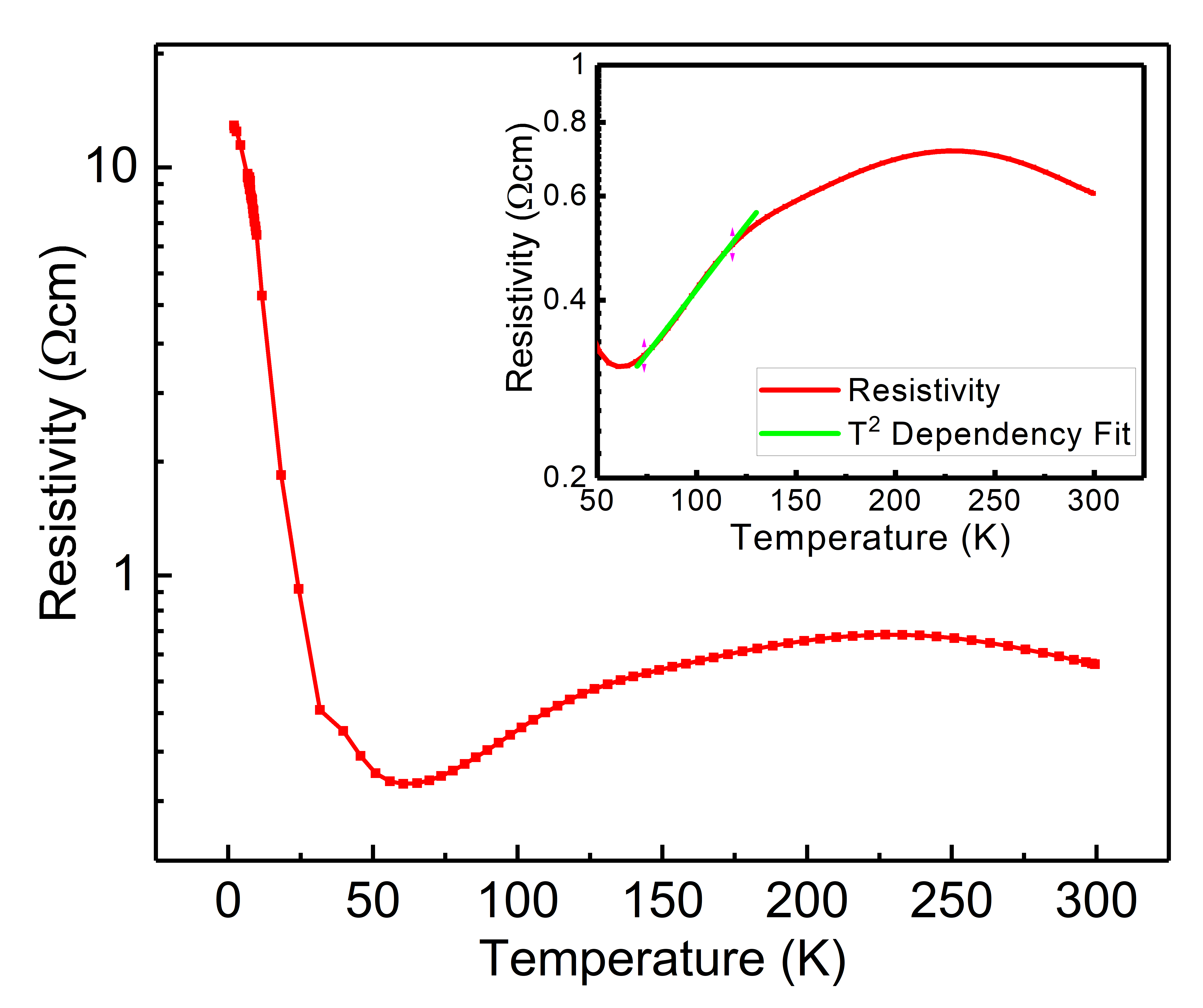}
\caption{\label{PLA} 
Temperature dependent resistance measurement for a representative La-rich film LVO-L3 measured with 0.001 mA. There are two turnovers in the full temperature range scan. Part of its metallic region can be fitted to Fermi-liquid model $\rho \sim$ T$^2$ (see inset).} 
\end{figure}

\begin{figure}[h]
\centering
\includegraphics[width=3.3in]{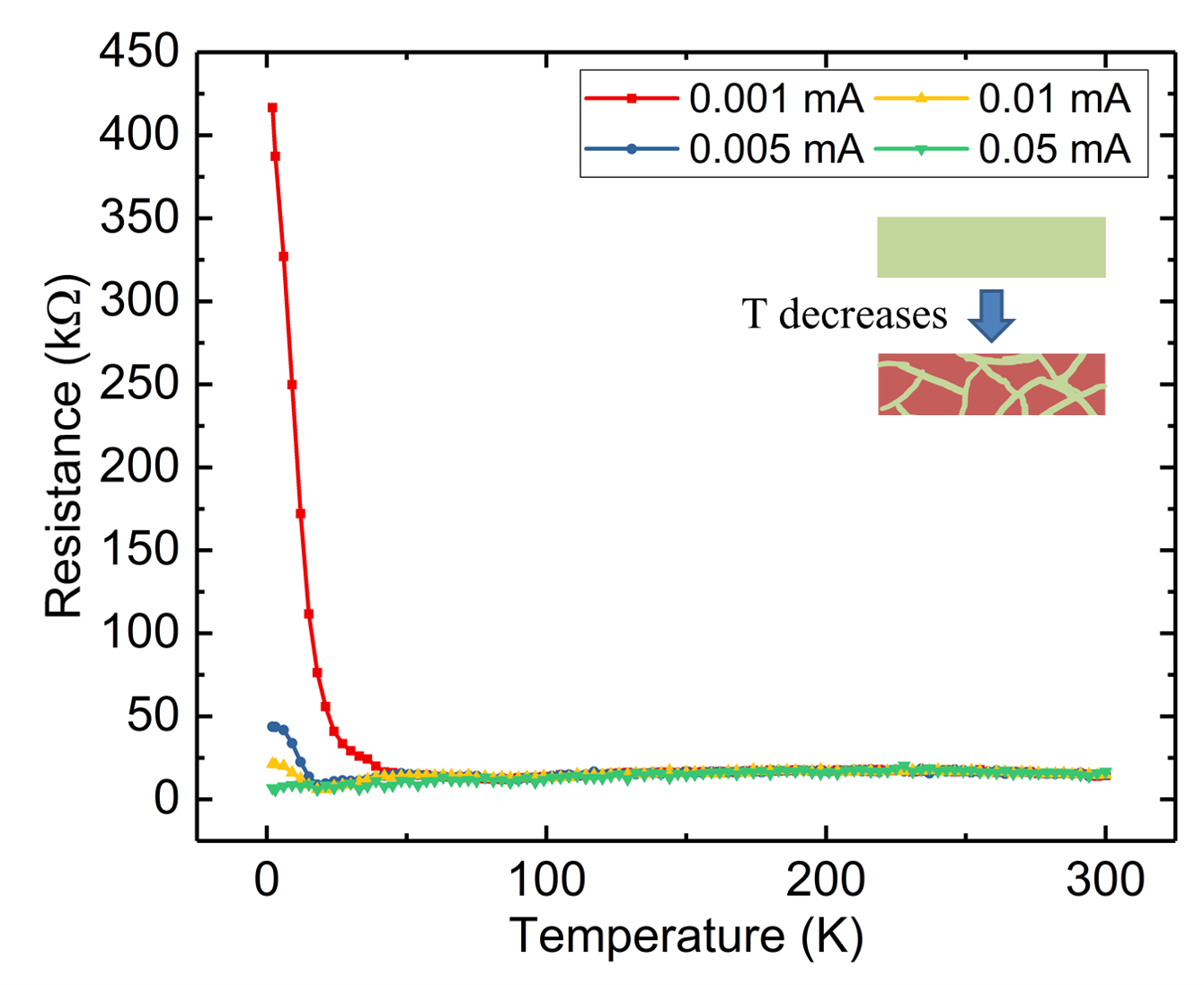}
\caption{\label{PI} The upturn in resistance at low T can be suppressed by increasing driving current; inset: a scheme showing insulating phase (colored in red) is not formed homogeneously allowing current to percolate when higher currents are applied (colored in green). }
\end{figure}

The situation is more complicated in La-rich films. Existing studies on bulk LVO show that samples become more insulating as they become more La-rich (or V-deficient)\cite{Gharetape2011EffectLaVO3}, but this is not what we observe in our films. In Fig.~\ref{PLA}, the temperature dependent resistivity of LVO-L3 shows two turnovers, the first one around 250~K is indicating that the film changes from an insulating-like state to a metallic-like state; the second one at low T, shows a drastic increase in resistance. Part of region between two turnovers can be fitted to $\rho \sim$ T$^2$, which indicates a Fermi-Liquid like behavior\cite{Gu2013Metal-insulatorFilms,Gu2013Metal-insulatorFilmsb}. All La-rich films, indicated by red symbols in Fig. \ref{XRD_c}, show this behavior (see Supplemental Materials \cite{suppmat} for more R vs. T data).

There have been reports on stoichiometric LVO thin films showing a similar temperature dependence of the resistivity. For example, He et al. reported the resistance of LVO changing from a high temperature conducting phase to a low T insulating phase in a 30 nm-thick film, attributing the observation to the LVO-STO interface effect\cite{PhysRevB.86.081401,Hotta2007PolarInterface}. But the LVO-STO interface likely does not affect our resistivity since our films are almost an order of magnitude thicker. Rather, it appears that the lattice parameter extension, and associated changes to the V-O-V bonding scheme, induced by V-deficiency is the crucial parameter controlling the transport properties. 
Additionally, the unexpected turnovers in our La-rich films could be due to the fact that V-vacancies cause the formation of defects and disorders inside the film, and lead to impurity-induced electron localization.
Thus, a phonon-related Peierls transition\cite{Peierls1955QuantumSolids} or disorder-related Anderson localization\cite{PhysRev.109.1492} can play a role in the observed turnovers in resistivity. 
Moreover, off-stoichiometry is often accompanied by the change in valence state. Since the film is La-rich, it is likely that the valence state of vanadium is not purely 3+. A mixed valence state can create a doping-like effect. Therefore the film behaves like a metal at high temperature because of excess charge carriers provided by the dopant, 
but at low temperature all electrons are bonded to chemical impurities and structural disorders so it becomes insulating again, similar to behaviors reported for Sr-doped LVO films\cite{Inaba1995ChangeLa1-xSrxVO3,doi:10.1063/1.3593489}. 

We also find that we can suppress the upturn in resistance at low T by increasing applied current (Fig.~\ref{PI}), i.e., the I-V curves are nonlinear at low temperature (see Supplemental Materials \cite{suppmat}). We can rule out heating as a source of nonlinearity in the low temperature I-V curves by performing a Joule heating analysis similar to the one reported in Padhan et al. \cite{PhysRevB.70.134403} (see Supplemental Materials for details). 
Razavi et al. reported similar upturns in resistivity and non-linear I-V characteristics at low temperature, they attributed it to the formation of random diode networks in their LVO thin films\cite{doi:10.1063/1.3293439}. They report that the observed nonlinear behavior happened only in the films with relatively small values for \textit{c} (close to that of the STO substrate) and it would disappear when the films become insulating as the lattice parameter of the film became larger (1$\%$ larger than that of STO). In their study the effect of possible off-stoichiometry was not discussed. All our La-rich films that have an out-of-plane lattice parameter of about 3.96 \AA~ or larger, show similar diode-like behavior. It is likely that the randomly distributed impurities and defects in the La-rich films help to make it conductor-like in the temperature range of 100~K to 250~K, but when the temperature is low enough they cause the localization of electrons resulting in scattered insulating regions. Once the randomly formed insulating regions become large enough to reach across the measured strip, the transport behavior will be dominated by the insulating state. However, given enough applied bias across this region, much like overcoming the contact potential in the forward bias mode of a metal-semiconductor junction, the current will start percolating through the strip and lower the measured resistance.

\section{Conclusion}
We have successfully grown off-stoichiometric LVO thin films on STO substrates using PLD technique by adjusting the laser fluence during growth. Clearly the off-stoichiometry in LaVO$_3$ films affects its structural, optical and transport performance. The absorption spectra, show little dependence on stoichiometric variation, however both types of films show absence of long-range orbital order, which is different from bulk. This is either due to the presence of structural domains in the films or it may be due to the off-stoichiometry distorting the V-O-V bonding scheme of the LVO films. 
While the V-rich films behave as Mott-insulators, the resistivity measurements 
in La-rich films show unexpected metal-insulator transitions and clear signatures of electronic phase separation at low temperature. It is clear that structural distortions due to the extension of the out-of-plane lattice parameter as well as the presence of doping and defects play a role in the changed transport properties of the La-rich films.  
The low temperature metal-insulator transition and the associated current-driven melting of electronic phase separation in La-rich films can make it useful in some temperature- or current- driven novel devices.  But when it comes to employing the material as an absorber in solar cells, one should adjust growth conditions so that the product tends to be V-rich rather than La-rich. However, the possible effects of charge-trapping due to defects associated with the observed anti-phase boundaries could hamper the use of these V-rich films in photovoltaic devices, which needs further investigation. 


A portion of this work was performed at the National High Magnetic Field Laboratory, which is supported by National Science Foundation Cooperative Agreements No. DMR-1157490 and No. DMR-1644779, and the State of Florida. C.B. acknowledges support from the National Research
Foundation, under grant NSF DMR-1847887. Use of the Advanced Photon Source was supported by the U.S. Department of Energy (DOE), Office of Science User Facility, operated for the DOE Office of Science by Argonne National Laboratory under Contract No. DE-AC02-06CH11357. Extraordinary facility operations were supported in part by the DOE Office of Science through the National Virtual Biotechnology Laboratory, a consortium of DOE national laboratories focused on the response to COVID-19, with funding provided by the Coronavirus CARES Act. S.M. acknowledges support under grant NSF DMR-1229217.

\bibliography{references}

\end{document}


\widetext
\begin{center}
\textbf{\Large{\textit{Supplemental Materials:} Growth and Characterization of Off-Stoichiometric LaVO$_3$ Thin Films}}
\end{center}

\setcounter{equation}{0}
\setcounter{figure}{0}
\setcounter{table}{0}
\setcounter{page}{1}
\makeatletter
\renewcommand{\theequation}{S\arabic{equation}}
\renewcommand{\thefigure}{S\arabic{figure}}
\renewcommand{\thetable}{S\arabic{table}}
\renewcommand{\thebibliography}{S\arabic{table}}
\renewcommand*{\citenumfont}[1]{S#1}
\renewcommand*{\bibnumfmt}[1]{[S#1]}


\subsection{Target preparation}
\begin{figure}[h]
\centering
\includegraphics[width=4.5in]{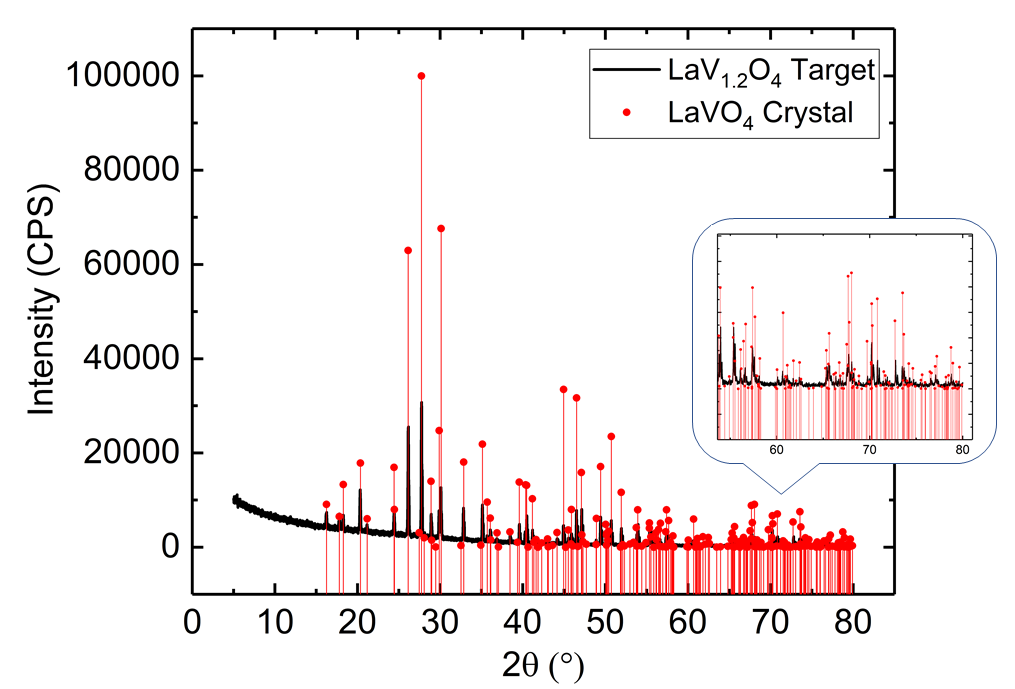}
\caption{\label{fig:XRD-Target}The XRD scan of the prepared target (black line) shows agreement with the theoretical spectrum of a bulk LaVO\textsubscript{4} crystal (red dot).}
\end{figure}

The LaV\textsubscript{1.2}O\textsubscript{4} target is prepared based on the existing recipe\cite{Kim2011Room}. The La\textsubscript{2}O\textsubscript{3} powder is pre-heated at 550 \textsuperscript{$\circ$}C to remove any impurity phases. The mixture of La\textsubscript{2}O\textsubscript{3} and V\textsubscript{2}O\textsubscript{5} powders is weighed following a 1:1.2 ratio, then manually ground and pressed into a pellet with an applied pressure of 20 MPa. Finally, the pressed pellet is sintered at 700 \textsuperscript{$\circ$}C and 1000 \textsuperscript{$\circ$}C in ambient for 12 hours each. After the process, we took the reacted pellet to perform XRD and confirmed the LaVO\textsubscript{4} phase (see Fig. \ref{fig:XRD-Target}).

Because the first baking temperature (for powders to react) is set to be 700\textsuperscript{$\circ$}C, higher than the melting point of V\textsubscript{2}O\textsubscript{5}, some vanadium is lost during the procedure. We prepared several targets with varying amounts of V\textsubscript{2}O\textsubscript{5} powder, while keeping the  quantity of La\textsubscript{2}O\textsubscript{3} powder the same. In this way we fabricated three different targets, LaVO\textsubscript{4}, LaV\textsubscript{1.1}O\textsubscript{4} and LaV\textsubscript{1.2}O\textsubscript{4} (targets are named after the starting ratio of the La\textsubscript{2}O\textsubscript{3} and V\textsubscript{2}O\textsubscript{5} powders). Based on a systematic study of thin film growth from each of these targets, we found that we can grow La rich/poor thin films out of the LaV\textsubscript{1.2}O\textsubscript{4} target based on laser fluence. Note, half of the La-rich films presented in the manuscript were grown using the LaV\textsubscript{1.1}O\textsubscript{4} target, the other half were grown using the LaV\textsubscript{1.2}O\textsubscript{4} target. All of the V-rich films were grown using the LaV\textsubscript{1.2}O\textsubscript{4} target. While the stoichiometry of the target is not the decisive factor to determine the La:V ratio of the film, starting with excess Vanadium in the target manufacturing procedure is required to get a close to stoichiometric target from which we are able to grow V-rich LVO films. The stoichiometry of our films is determined from EDS after the growth.

\subsection{Additional X-Ray Diffraction information on thin films}

\begin{figure}[h]
\centering
\includegraphics[width=7in]{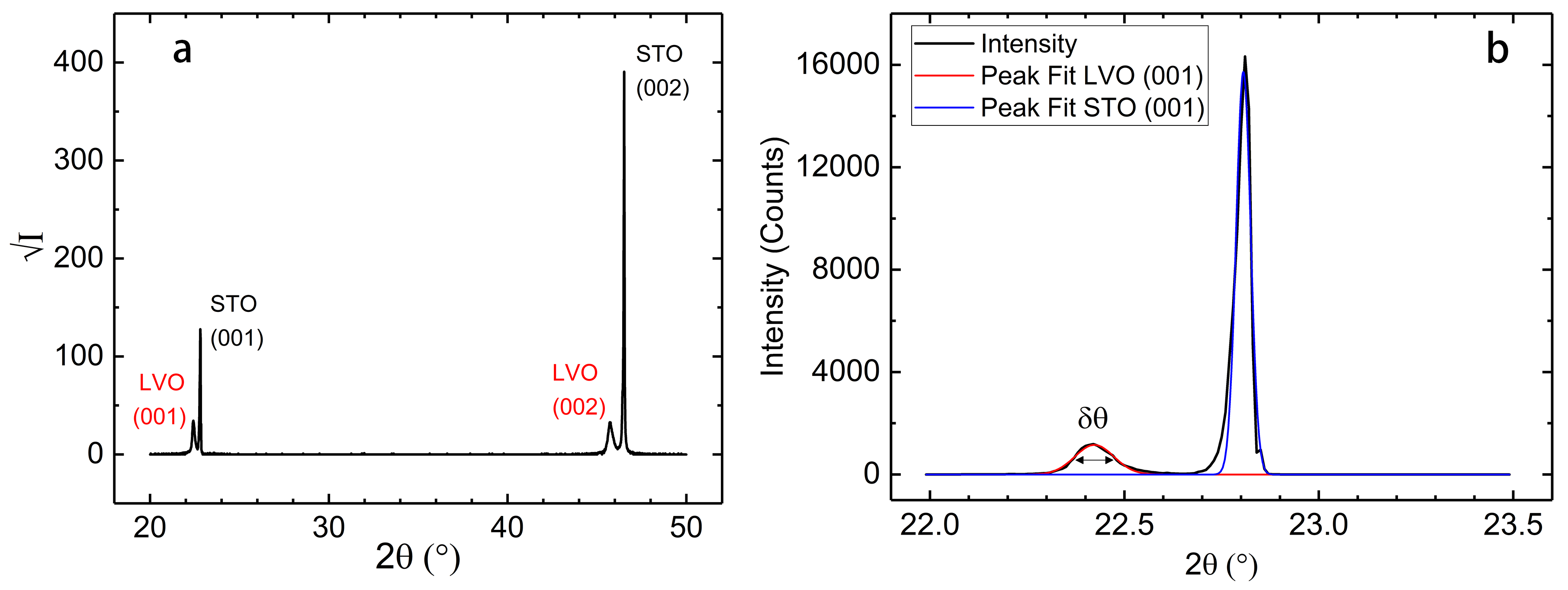}
\caption{\label{fig:XRD-supp}(a) A $\theta-2\theta$ scan through the (00L) peaks of a representative LVO film. The y-axis is the square root of intensity. Both film and substrate peaks are labeled. (b) The Gaussian fit (red line) of LVO (001) gives the value of $\theta$. We use the value extracted from FWHM ($\delta\theta$, indicated by a double arrow) to determine the error bar $\delta\textit{c}$.}
\end{figure}

LVO films' out-of-plane lattice constants \textit{c} are determined based on Bragg's Law:
\begin{equation}
    2dsin\theta = n \lambda
\end{equation}

$\theta$ is the peak position extracted from Gaussian fit of LVO (00L) peak (Fig.~\ref{fig:XRD-supp}). $\delta\theta$ is extracted from the FWHM value of (00L) peak. The error bar $\delta\textit{c}$ is determined using the principle of error propagation:
\begin{eqnarray}
\begin{array}{cc}
   d(\theta+\delta\theta) \approx d(\theta) + \delta d(\theta)\\\\
  \delta d (\theta) = d'(\theta)\delta\theta
\end{array}
\end{eqnarray}

The calculated out-of-plane lattice constant for all our LVO films are plotted in figure 4 in the main text. Overall, V-rich films have smaller out-of-plane constant than La-rich films.


To calculate the lattice constants from RSM, we use OriginLab to extract the intensity profile around the peaks; then fit the line profile to Lorentz function to find the peak position \textit{q}. Once we have the \textit{q} coordinates for the LVO film, we use the following equation to calculate all three lattice constants:

\begin{equation}
   a_{LVO} = a_{STO} \frac{q_{STO}}{q_{LVO}}
\end{equation}

\begin{figure}[h]
\centering
\includegraphics[width=7in]{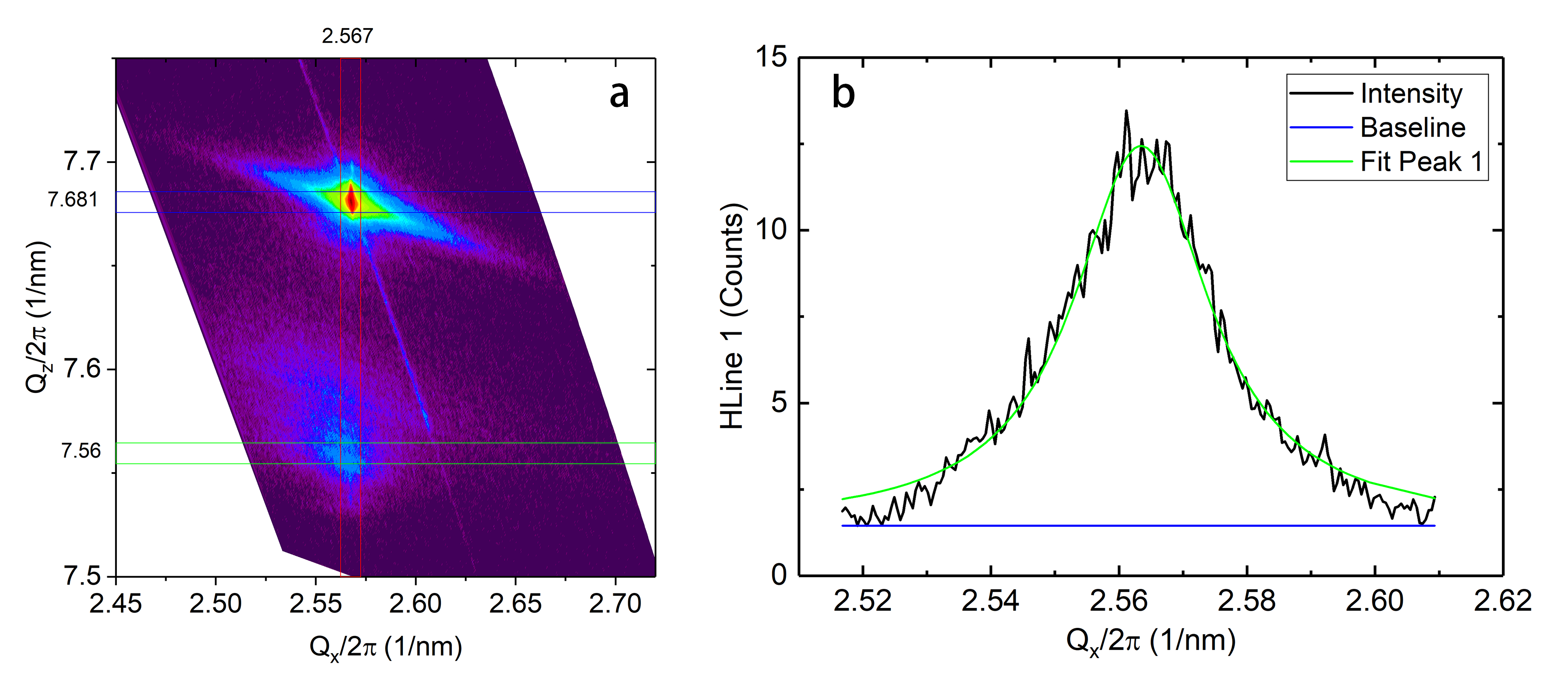}
\caption{\label{fig:RSM-Supp}(a) An example of extracting line profiles from the LVO-L1 (103) RSM plot. (b) Lorentz fit to the line profile of the film peak at Q\textsubscript{z}/2$\pi$ = 7.56 nm\textsuperscript{-1}.}
\end{figure}
where \textit{a\textsubscript{STO}} = 3.905\AA. The error bars \textit{$\delta$a} are calculated using the same error propagation principle:

\begin{equation}
     \delta a (q) = a'(q)\delta q
\end{equation}
\textit{$\delta$q} is the standard deviation of the peak position from Lorentz fit.




\subsection{Additional transport plots on thin films}

\begin{figure}[b]
\centering
\includegraphics[width=4in]{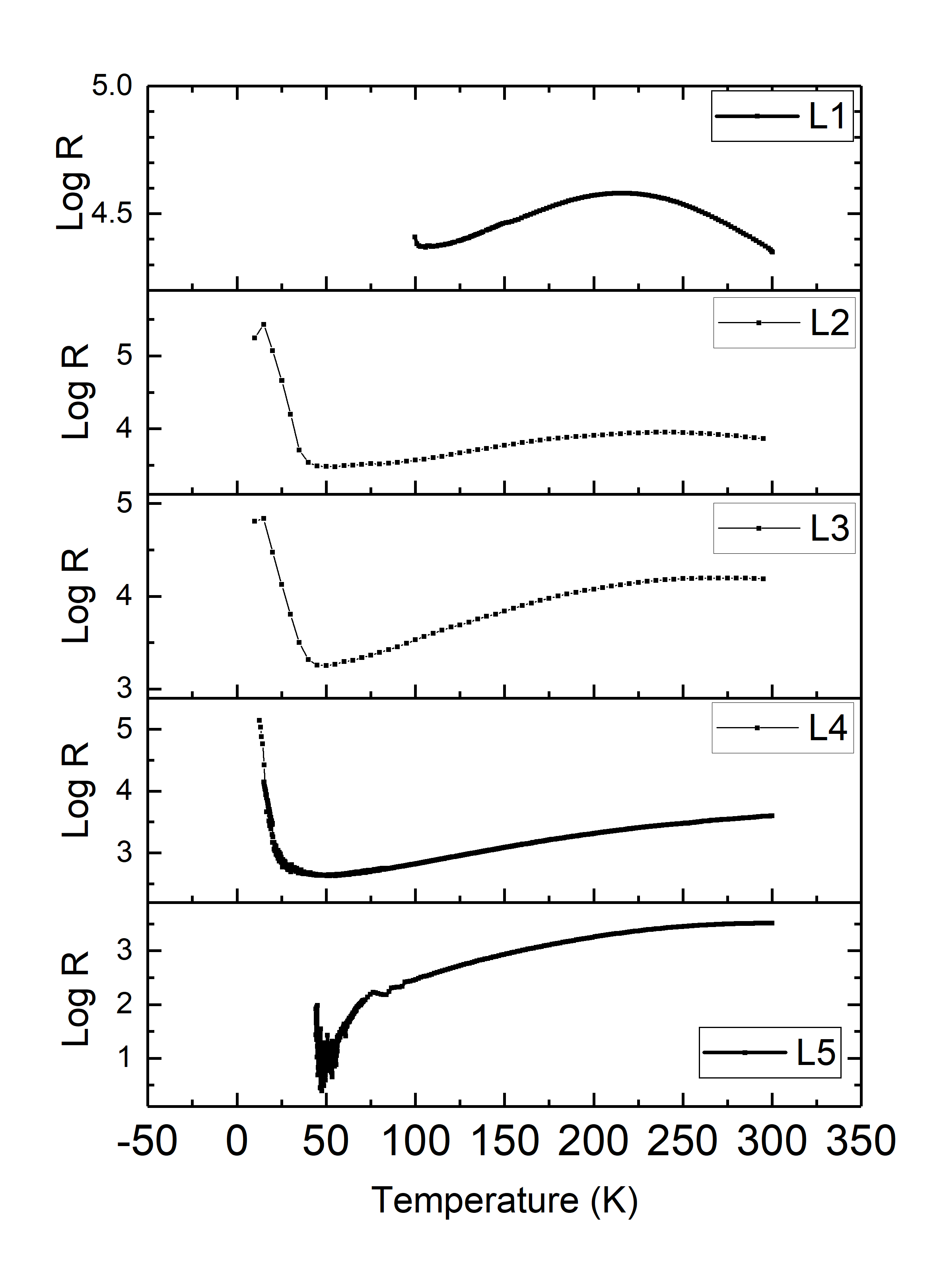}
\caption{\label{fig:RTall}R vs T measurements on selective La-rich samples. In this set of measurements, the contacts were made by cold welding indium method and the samples are not cut into strips.}
\end{figure}

\begin{figure}[h]
\centering
\includegraphics[width=5in]{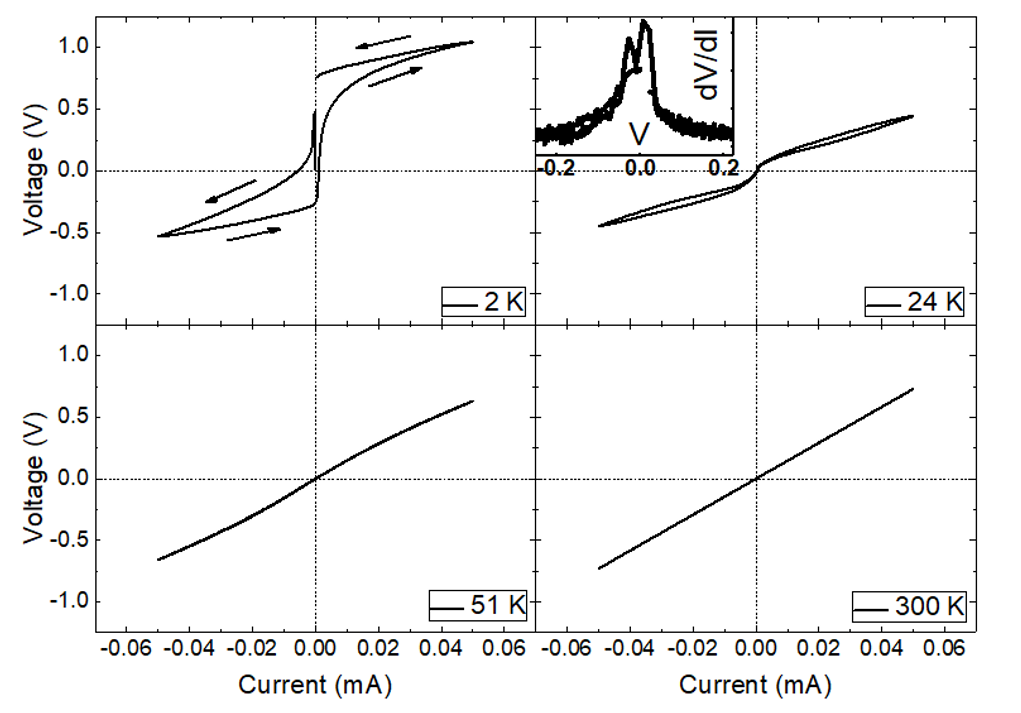}
\caption{\label{fig:IV-La}I-V characteristic of LVO-L3. The I-V curve starts to deviate from linear when then temperature is below $\sim$50~K. The inset show the derivative dV/dI vs. V taken for the 24~K IV to highlight that the main resistance drop occurs around 0.1~V. For the IV at 2~K, the arrows indicate the direction of the current sweep. }
\end{figure}

\begin{figure}[h]
\centering
\includegraphics[width=4in]{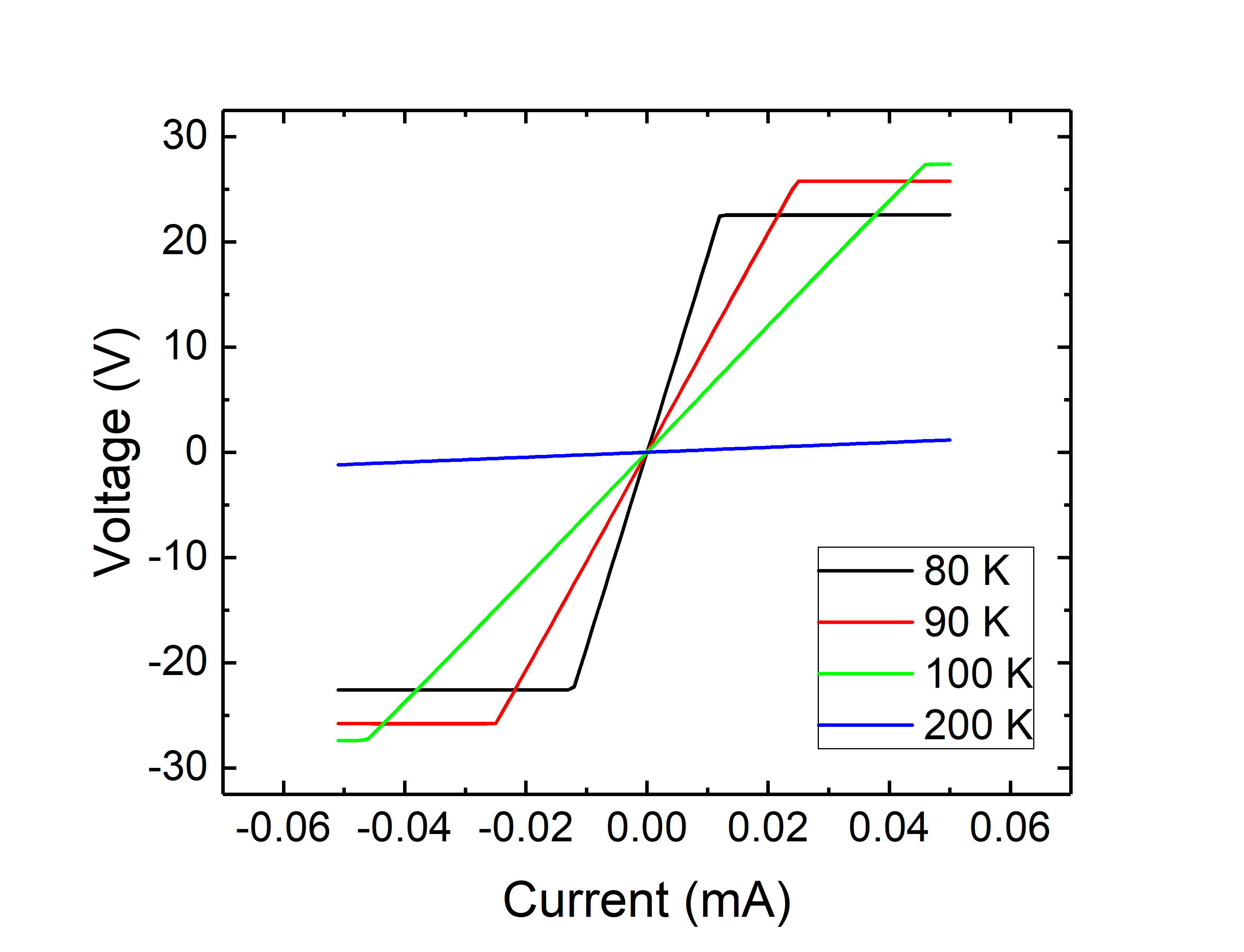}
\caption{\label{fig:IV-V}I-V characteristic of LVO-V2. All the I-V curves remain linear at all temperatures. The apparent saturation is an artefact due to compliance and/ or range limits of the current source and volt meter.}
\end{figure}

R vs.T measurements for various La-rich films are shown in Fig. \ref{fig:RTall} showing the reproducibility of the various T-dependent turnovers in the transport properties. 
Representative IV characteristics for La and V rich samples are shown in Figs. \ref{fig:IV-La} and \ref{fig:IV-V}. 

\begin{figure}
\centering
\includegraphics[width=7in]{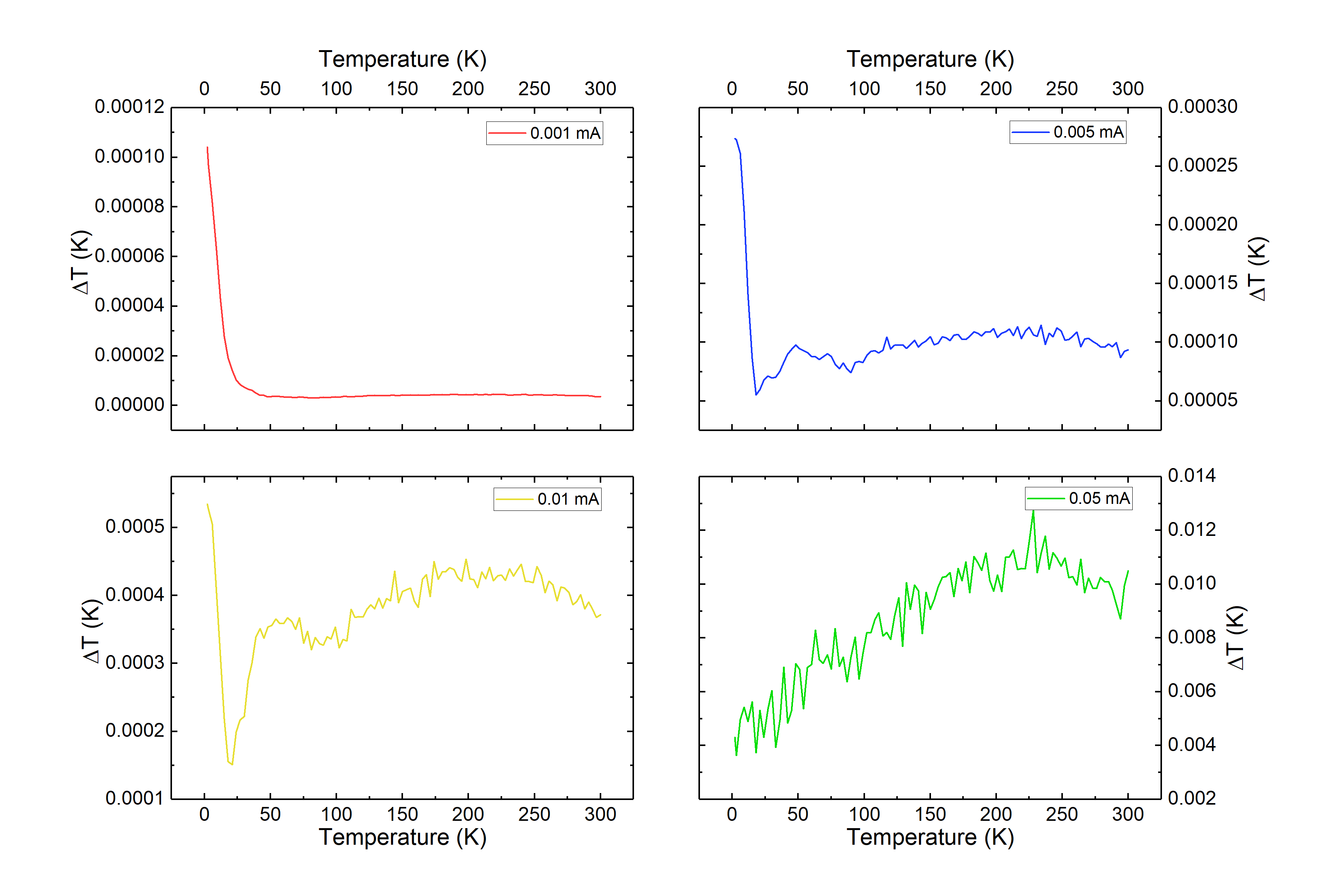}
\caption{\label{fig:dT}Estimation of temperature change $\Delta$T due to Joule heating.}
\end{figure}

To estimate the heating cause by applied current, we use the equation to calculate the temperature increase $\Delta$T: \cite{PhysRevB.70.134403}

\begin{equation}
   \Delta T = \frac{2I^2\rho (T+\Delta T)} {S\kappa_{sub} (T+\Delta T)}
\end{equation}
with I the applied current, $\rho$ the resistivity, $S$ the cross section of the measured strip, and $\kappa$ the thermal conductivity of the substrate. 
The $\Delta$T vs T plots at different currents are in Fig.\ref{fig:dT}. These calculations are based on the resistance vs. temperature curves presented in Fig. 9 in the main text, using $L$ = 0.5x10$^{-3}$ m (the spacing between gold contacts), $S$ = 0.6x10$^{-3}$~x~350x10$^{-9}$~m$^2$ and $\kappa$ = 16 WK$^{-1}$m$^{-1}$.
\bibliography{references}